\documentclass[journal]{IEEEtran}
\usepackage{graphicx}
\usepackage{epstopdf}
\usepackage{cite}
\usepackage{algorithm}
\usepackage{algorithmic}
\usepackage{amsmath}
\usepackage{bm,comment,color,amssymb}
\usepackage{stfloats}
\usepackage{caption}
\usepackage{subcaption}

\usepackage[colorlinks]{hyperref}

\begin{document}

\title{ Paving the Way to 6G: Outage Probability \\Analysis for FAS-ARIS Systems }

\author{Jianchao Zheng, Xiazhi Lai, Junteng Yao, Jie Tang, \\  Yijin Pan, Tuo Wu, Chau Yuen, \emph{Fellow, IEEE} 

\thanks{ \emph{(Corresponding authors: Tuo Wu)}.
}
\thanks{ J. Zheng is with the School of Computer Science and Engineering, Huizhou University, Huizhou 516007, China (E-mail: $\rm zhengjch@hzu.edu.cn.$). X. Lai is with the School of Computer Science, Guangdong University of Education, Guangzhou, Guangdong, China (E-mail: $\rm xzlai@outlook.com$). J. Yao is with the Faculty of Electrical Engineering and Computer Science, Ningbo University, Ningbo 315211, China (E-mail:  $\rm yaojunteng@nbu.edu.cn$). J. Tang is with the School of Electronic and Information Engineering, South China University of Technology, Guangzhou 510640, China (E-mail: $\rm eejtang@scut.edu.cn$). T. Wu and C. Yuen are with the School of Electrical and Electronic Engineering, Nanyang Technological University, 639798, Singapore (E-mail: $\rm \{tuo.wu, chau.yuen\}@ntu.edu.sg$). Y. Pan is with National Mobile Communications Research Laboratory, Southeast University, Nanjing 	210096, China. (E-mail: $\rm panyj@seu.edu.cn$).
}
} 

\markboth{}
{Zheng \MakeLowercase{\textit{et al.}}:  FAS-ARIS Communication Systems Analysis with Block-Correlation Model}

\maketitle

\begin{abstract}
In this paper, we pave the way to six-generation (6G) by investigating the outage probability (OP) of fluid antenna system (FAS)-active reconfigurable intelligent surface (ARIS) communication systems. We consider a FAS-ARIS setup consisting of a base station (BS) with a single fixed-position antenna and a receiver equipped with a fluid antenna (FA). Utilizing the block-correlation model, we derive a closed-form expression for the OP. Our analysis, supported by numerical results, confirms the accuracy and effectiveness of the derivation. Furthermore, the results demonstrate that the FAS-ARIS system significantly outperforms other configurations in terms of OP, highlighting its potential to enhance communication performance and reliability in future 6G networks.
\end{abstract}

\begin{IEEEkeywords}
Fluid antenna system (FAS),  active reconfigurable intelligent surfaces (ARIS), outage probability (OP).
\end{IEEEkeywords}
\IEEEpeerreviewmaketitle

\section{Introduction}
\IEEEPARstart{T}{he} six-generation (6G) wireless communication system requires higher data rates to accommodate the diverse devices in the internet of things (IoT) networks. However, traditional wireless massive multiple input multiple output (MIMO) technology faces increasing costs with additional antennas, making fluid antenna system (FAS) a promising alternative \cite{FAS20}. Practical implementations of FAS include pixel-based structures \cite{Rodrigo14} and liquid metal \cite{Huang21}.  FAS has become a highly active research topic, as evidenced by recent studies \cite{XLai23, JZheng24, YaoJ24, HXu24}. 

In FAS-assisted communication systems, optimizing the positioning of fluid antennas (FAs) is crucial for selecting the best channels and maximizing performance \cite{Alvim-2023, Dai-2023, Xu-2024}, although this process can be complex and costly. Reconfigurable intelligent surfaces (RIS) \cite{Khoshafa-2021, TWu1, TWu2} offer a solution by adjusting the phase of reflected signals, thereby compensating for suboptimal FA positioning and mitigating performance losses. Due to these benefits, FAS-RIS systems have gained significant research interest, with studies focusing on outage performance \cite{LaiX242, Ghadi2024} and optimization \cite{YaoJ242, YaoJ243}.

However, RIS deployment often results in limited performance gains due to the ``multiplicative fading" effect, which leads to higher path loss compared to direct links \cite{KZhi221}. To address these limitations, Yao \textit{et al.} proposed replacing passive RIS with active RIS (ARIS) for FAS-RIS systems and investigated its optimization in \cite{YaoJ243}. ARIS employs active reflective amplifiers to amplify the reflected signals, resulting in improved performance compared to passive RIS. 

ARIS employs active reflective amplifiers to amplify the reflected signals, resulting in improved performance compared to passive RIS \cite{Dai2}. However, the performance of FAS-ARIS systems is not well understood and remains unexplored, particularly in the context of paving the way to 6G networks. Motivated by this gap, this paper investigates a FAS-ARIS system, considering both the reflected link between the FAS-ARIS-base station (BS) and the direct link between the FAS and BS. By applying the block-correlation (BC) model, we derive approximate expressions for the outage probability (OP). Simulation results demonstrate that the FAS-ARIS system significantly outperforms other configurations, highlighting its potential to enhance communication performance and reliability in future 6G networks.
\vspace{-5pt}
\section{System Model} 
Consider a FAS-ARIS communication system consisting of a base station (BS) equipped with a single fixed-position antenna (FPA), a receiver featuring an FA  capable of switching among \(N\) ports within a linear space of \(W\lambda\), and an ARIS composed of \(M\) reflecting elements. In this configuration, the BS transmits signals to the receiver through both direct channels and ARIS-assisted pathways. 
  
\subsection{Communication Model}
The BS transmits the signal to receiver through the direct link and reflections from the ARIS simultaneously. Accordingly, the signal received at the $k$-th FAS port of the receiver is represented by
\begin{align}
y_k=&\sqrt{P}s\left( \sum_{m=1}^M h_m \omega v_{m,k} e^{-2i\pi\theta_m}\left(\frac{d_{sr}d_{rd}}{d_0}\right)^{-\frac{\alpha}{2}}\right.\nonumber\\
&\left.+\chi_k \left(\frac{d_{sd}}{d_0}\right)^{-\frac{\alpha}{2}}\right)+\omega n_k+n_r,
\end{align}
where \(\omega\), \(d_0\), and \(\alpha\) represent the amplification factor of each RIS element, the reference distance, and the path loss exponent, respectively. The system assumes Rayleigh fading channels, which are appropriate for non-line-of-sight conditions. Specifically, \(h_m \sim \mathcal{CN}(0, \epsilon_1)\) represents the channel coefficient between the BS and the \(m\)-th RIS element, while \(v_{m,k} \sim \mathcal{CN}(0, \epsilon_2)\) denotes the channel coefficient from the \(m\)-th RIS element to the \(k\)-th port of the receiver. Additionally, \(\chi_k \sim \mathcal{CN}(0, \epsilon_3)\) represents the channel coefficient from the BS directly to the \(k\)-th port. The transmitted signal is denoted by \(s \sim \mathcal{CN}(0, 1)\) with transmission power \(P\). Besides, \(n_k \sim \mathcal{CN}(0, \sigma_k^2)\) and \(n_r \sim \mathcal{CN}(0, \sigma_r^2)\) represent the additive white Gaussian noise (AWGN) for the \(k\)-th port and the receiver, respectively. The reflection phase shift of the \(m\)-th RIS element is denoted by \(\theta_m\), and the RIS optimizes the channel gain by setting \(\theta_m = -\text{arg}(h_m v_{m,k})\) when the phases of \(h_m\) and \(v_{m,k}\) are known. Consequently, the channel gain for the \(k\)-th  port at the receiver is given by\vspace{-5pt}
\begin{align}  \label{aa1}
	\gamma_k= \sum_{m=1}^M|h_m||v_{m,k}|.
\end{align}
\vspace{-2pt}
Given that $|h_m|$, and $|v_{m,k}|$ follow a Rayleigh distribution, the expected value and variance of their product can be derived as $\mathbf{E}\left(|h_m||v_{m,k}|\right)=\mathbf{E}\left(|h_m|\right)\mathbf{E}\left(|v_{m,k}|\right)=\pi\sqrt{\epsilon_1\epsilon_2}/4$, $\mathbf{Var}\left(|h_m||v_{m,k}|\right)=\mathbf{E}\left(|h_m|^2|v_{m,k}|^2\right)-\left(\mathbf{E}\left(|h_m||v_{m,k}|\right)\right)^2=\epsilon_1\epsilon_2(1-\pi^2/16)$. Besides, the maximum channel gain can be obtained by selecting the optimal port for signal receiving. Hence, the channel parameter of the optimal port is given by
\begin{align}
	\gamma^*=\mathop{\text{max}}\limits_{k=1,\cdots, N} \gamma_k.
\end{align}
Consequently, the signal-to-noise ratio (SNR) is given by
\begin{align}
\gamma=\left|\Omega_1 \gamma^*+\Omega_2 |\chi| \right|^2,
\end{align}
and 
\begin{align}
	\Omega_1= \sqrt{\frac{P}{\sigma^2}}\left(\frac{d_{sr}d_{rd}}{d_0}\right)^{-\frac{\alpha}{2}}, 
	\Omega_2=  \sqrt{\frac{P}{\sigma^2}}\left(\frac{d_{sd}}{d_0}\right)^{-\frac{\alpha}{2}},
\end{align}
where $\sigma^2=\eta^2\sigma_k^2+\sigma_r^2$, and  $|\chi|$ denotes the channel parameter from the BS to the receiver based on a qualified FAS port, and it follows the Rayleigh distribution with $\epsilon_3$. Hence, the probability density function (PDF) of $|\chi|$ can be formulated as
\begin{align}
f_{|\chi|}(x)=\frac{x}{\epsilon_3}\exp\left(\frac{x^2}{2\epsilon_3^2}\right),
\end{align} 
\subsection{FAS Channel Correlation Model}
For the channel correlation between distinct FAS ports, the 3D Clarke's model \cite{Espinosa24} is applied. The correlation coefficient between ports $\varsigma$ and $\tau$ can be modeled as 
\begin{align}
	g(\varsigma,\tau)=\text{sinc}\left(\frac{2\pi (\varsigma-\tau) W}{N-1}\right),
\end{align}
where $\mathrm{sinc}(x)=\frac{\sin(x)}{x}$. Since $\mathrm{sinc}(x)$ is an odd function, the correlation coefficient matrix takes the form of a Toeplitz matrix, which can be represented as
\begin{equation}
	\mathbf{\Sigma}\in\mathbb{R}^{N\times N}=
	\begin{bmatrix}\label{q5}
		g_{1,1} & g_{1,2} & \dots & g_{1,N}\\
		g_{1,2} & g_{1,1} & \dots & g_{1,N-1}\\
		\vdots &  \ddots & \vdots \\
		g_{1,N} & g_{1,N-1} & \dots & g_{1,1}
	\end{bmatrix}.
\end{equation}

\section{Outage Performance analysis}
In this section, we analyze the outage performance of the FAS-ARIS system. Let $R$ represent the target transmission rate, and the OP can be expressed as
\begin{align}
	P_{\mbox{out}}=F_{\gamma}\left(\beta\right)=\Pr\left(\log_2\left(1+\gamma\right)<R\right),
\end{align}
where $\beta=2^R-1$. To evaluate the outage probability, it is essential to derive the cumulative distribution function (CDF) of $\gamma$, i.e. $F_{\gamma}\left(y\right)$. Since obtaining the exact form of $F_{\gamma}\left(y\right)$ is challenging, the following subsections will present several approximations for $F_{\gamma}\left(y\right)$. These results are developed based on the central limit theorem (CLT).

\subsection{Block-Correlation Channel Approximation}
To efficiently capture the dominant eigenvalues, we apply the BC approximation model \cite{Espinosa24} to simplify the form of Eq. \eqref{q5}. As a result, the correlation coefficient matrix can be rewritten as
\begin{equation}
  \hat{ \mathbf{\Sigma}}\in\mathbb{R}^{N\times N}=
	\begin{bmatrix}
	\mathbf{A}_1 &\mathbf{0} &\mathbf{0} & \dots &\mathbf{0}\\
	\mathbf{0} &\mathbf{A}_2 &\mathbf{0}& \dots & \mathbf{0}\\
	 \vdots &  \ddots & \vdots \\
	 \mathbf{0}& \mathbf{0} & \mathbf{0}& \dots & \mathbf{A}_B
	\end{bmatrix},
\end{equation}
where 
\begin{equation}
 \mathbf{A}_b\in\mathbb{R}^{L_b\times L_b}=
	\begin{bmatrix}
	1 &\mu &\mu & \dots &\mu\\
	\mu &1&\mu& \dots &\mu\\
	 \vdots &  \ddots & \vdots \\
	 \mu& \mu & \mu& \dots & 1
	\end{bmatrix},
\end{equation}  
Here, $\mu$ is a value close to 1, and $\sum_{L_b=1}^B L_b=N$, where $B=|\it{S(\lambda_{th})}|$ represents the cardinality of the set  $\it{S(\lambda_{th})}$. Further, $\it{S(\lambda_{th})}=\left\{\lambda_n|\lambda_n\geq\lambda_{th}, n=1,\cdots,N\right\}$, where $\lambda_{th}$ is a small number that ensures a sufficient number of eigenvalues are included and can be dynamically adjusted. The value of $L_b$ is determined by the following criterion
\begin{align}
\mathop\text{arg\ min}\limits_{L_1,\cdots, L_B} \text{dist}( \mathbf{\Sigma}, \hat{ \mathbf{\Sigma}}),
\end{align}
where dist(·) is a distance metric between two matrices, which is determined by the difference of their eigenvalues and the detailed procedure can be found in \cite{Espinosa24}. 

Based on the BC model described above, and assuming that the number of RIS elements $M$ is sufficiently large, the CLT approximation can be applied \cite{LaiX242}.  Under this approximation, $\mathbf{\Gamma}=[\gamma_1,\cdots,\gamma_N]$ can be approximated as $\hat{\mathbf{\Gamma}}=[\hat{\gamma}_1,\cdots,\hat{\gamma}_N]$. Consequently, applying the Pearson correlation coefficient calculation, the correlation coefficient matrix of $\hat{\mathbf{\Gamma}}$ is expressed as
\begin{equation} 
	\label{bn}
	\bar{ \mathbf{\Sigma}}\in\mathbb{R}^{N\times N}=
	\begin{bmatrix}
		\mathbf{D}_1 &\mathbf{C} &\mathbf{C} & \dots &\mathbf{C}\\
		\mathbf{C} &\mathbf{D}_2 &\mathbf{C}& \dots & \mathbf{C}\\
		\vdots &  \ddots & \vdots \\
		\mathbf{C}&\mathbf{C} & \mathbf{C}& \dots & \mathbf{D}_B
	\end{bmatrix},
\end{equation}
where 
\begin{equation}
	{ \mathbf{D}_b}\in\mathbb{R}^{N\times N}=
	\begin{bmatrix}
		1 &\eta(\mu) &\eta(\mu) & \dots &\eta(\mu)\\
		\eta(\mu) &1 &\eta(\mu)& \dots & \eta(\mu)\\
		\vdots &  \ddots & \vdots \\
		\eta(\mu)&\eta(\mu)&\eta(\mu)& \dots & 1
	\end{bmatrix},
\end{equation}
Additionally, all elements in the submatrix of $\mathbf{C}$ are represented  as $\eta(0)$, while  $\eta(\mu_{k.l})$ is expressed as
\begin{align}
\eta(\mu_{k,l}) =\frac{M\epsilon_1 \mathbf{E}(|v_{m,k}||v_{m,l}|)-E_{\gamma}^2/M}{V_{\gamma}},
\end{align}
with 
\begin{align}
\mathbf{E}(|v_{m,k}||v_{m,l}|)=\int_0^\infty\int_0^\infty xy f_{|v_{m,k}|,|v_{m,l}|}(x,y)dxdy,
\end{align} 
and
\begin{align}
f_{|v_{m,k}|,|v_{m,l}|}(x,y)=\frac{4x^2y^2}{\epsilon_2^2(1-\mu_{k,l})}&e^{-\frac{x^2}{\epsilon_2}}e^{-\frac{y^2+\mu_{k,l}x^2}{\epsilon_2^2(1-\mu_{k,l})}}\nonumber\\
\times&I_0\left(\frac{2\mu_{k,l}xy}{\epsilon_2(1-\mu_{k,l})}\right),
\end{align}
where  $f_{|v_{m,k}|,|v_{m,l}|}(x,y)$ is the joint PDF of random variables (RVs) $|v_{m,k}|$ and $|v_{m,l}|$, $I_0(\cdot)$ denotes the zero order of Bessel function of the first kind. The term $\mathbf{E}(|v_{m,k}||v_{m,l}|)$ can be computed through numerical integration. Additionally, we have
\begin{align}
E_{\gamma}=&\frac{M\pi\sqrt{\epsilon_1\epsilon_2}}{4},\\
V_{\gamma}=&M\epsilon_1\epsilon_2\left(1-\frac{\pi^2}{16}\right).
\end{align}
Based on the structure of  $\bar{ \mathbf{\Sigma}}$ in Eq. \eqref{bn}, which shows that the elements of $\hat{\mathbf{\Gamma}}$ can be partitioned into $B$ parts, each containing $L_b$ identically distributed RVs,
for $b = 1, \cdots,  B$, we can then rearrange the subscripts and rewrite $\hat{\mathbf{\Gamma}}$ as  $[\hat{\gamma}_{1,1}, \cdots, \hat{\gamma}_{L_1,1}, \cdots, \hat{\gamma}_{k,d}, \cdots, \hat{\gamma}_{L_B,B}]$.  Here, $ \hat{\gamma}_{k,d}$ can be expressed as
\begin{align}
\hat{\gamma}_{k,b}=\sqrt{1-\rho_1}z_{2,k,d}+\sqrt{\rho_1-\rho_0}z_{1,k}+\sqrt{\rho_0}z_0+E_{\gamma},
\end{align}
for $k=1,\cdots, L_b$, $b=1,\cdots,B$, where $z_{2,k,d}$, $z_{1,k}$ and $z_0$ are independent and identically distributed (i.i.d.) Gaussian RVs with mean 0 and variance $V_{\gamma}$. Hence, the CDF of $\hat{\gamma}^*$ can be written as
\begin{align}
F_{\hat{\gamma}^*}(y)=&\frac{H\pi}{U^2 V_{\gamma}}\sum_{l=1}^U\prod_{b=1}^B\sum_{t_b=1}^U\frac{\left[1+\mbox{erf}\left(\frac{y-H p_{t,b}-H q_{l}}{s\pi V_{\gamma}(1-\rho_1)}\right)\right]^{L_b}}{2^{L_b+1}}\nonumber\\
&\times \sqrt{\frac{(1-p_{t,b}^2)(1-q_l^2)}{\rho_0(\rho_1-\rho_0)}}e^{-\frac{(H p_{t,b})^2}{2V_{\gamma}(\rho_1-\rho_0)}-\frac{(H q_l-E_{\gamma})^2}{2V_{\gamma}\rho_0}},
\end{align}
where $H$ is a large constant, $U$ is a parameter that controls the trade-off between accuracy and complexity, $p_{t,b}=\cos\left(\frac{(2t_{b}-1)\pi}{2U}\right)$, and $q_l=\cos\left(\frac{(2l-1)\pi}{2U}\right)$.

Furthermore, the CDF of $\gamma$ can be computed as
\begin{align}\label{q23}
F_{\gamma}(t)=\Pr(\gamma<t)=\Pr(\left|\Omega_1 \gamma^*+\Omega_2 |g|\right|^2<t),
\end{align}
after some mathematical manipulation, we obtain
\begin{align}
F_{\gamma}(t)=&\int_0^\infty F_{\hat{\gamma}^*}\left(\frac{\sqrt{t}-\Omega_2 x}{\Omega_1}\right)f_{|\chi|}(x)dx\nonumber\\
=&\frac{H\pi}{2U}\sum_{d=1}^U\sqrt{1-q_d^2}F_{{\hat{\gamma}^*}(y)}\left(\frac{\sqrt{t}-\Omega_2 \zeta}{\Omega_1}\right)f_{|\chi|}(\zeta)
\end{align}
where the Gauss-Chebyshev integral is applied in the final step, $q_d=\cos\left(\frac{(2d-1)\pi}{2U}\right)$, $\zeta=\frac{Hq_d+H}{2}$. Thus, the OP under BC approximation can be expressed as
\begin{align}
	P_{\mbox{out}}\approx\frac{H\pi}{2U}\sum_{d=1}^U\sqrt{1-q_d^2}F_{{\hat{\gamma}^*}(y)}\left(\frac{\sqrt{\beta}-\Omega_2 \zeta}{\Omega_1}\right)f_{|\chi|}(\zeta).
\end{align}

\subsection{CLT and I.I.D. Channel Approximation}
To gain further insight, we consider the case where $\mu=1$, i.e., perfect correlation between the ports within each block. In this scenario, each block can be treated as a single antenna, and the FAS system is directly approximated as a collection of $B$ independent antennas. Thus, we have
\begin{equation}
	\mathbf{D}_b\in\mathbb{R}^{L_b\times L_b}=\mathbf{1}_{N\times N}, 
\end{equation} 

Thanks to the BC model, where the correlation within each block remains constant, the analysis for FAS-RIS communication becomes manageable. Specifically, the expression for $\bar{\gamma}_k$ is given by
\begin{align}
\bar{\gamma}_k=\sqrt{1-\rho_0}d_k+\sqrt{\rho_0}d_0+E_{\gamma},
\end{align}
where $d_k\sim \mathcal{N}(0, V_{\gamma})$ for $k=0,1,2,\cdots, B$.   
Based on this, the CDF of $\bar{\gamma}^*$ can be formulated as 
\begin{align}
F_{\bar{\gamma}^*}(y)\approx  &\frac{H\pi}{U_l}\sum_{l=1}^U \frac{1}{2^B}\left[1+\text{erf}\left(\frac{y-E_{\gamma}-\rho_0\tau}{\sqrt{2V_{\gamma}(1-\rho_0)}}\right)\right]^B\nonumber\\
&\times\sqrt{\frac{1-q_l^2}{2\pi V_{\gamma}}}e^{-\frac{\tau^2}{2V_{\gamma}}},
\end{align}
where $\tau=\frac{Hq_l+H}{2}$. Following the same derivation as in Eq. \eqref{q23}, the CDF of $\gamma$ can be expressed as
\begin{align}\label{q23}
F_{\gamma}(t)=\frac{H\pi}{2U}\sum_{d=1}^U\sqrt{1-q_d^2}F_{{\bar{\gamma}^*}(y)}\left(\frac{\sqrt{t}-\Omega_2 \zeta}{\Omega_1}\right)f_{|g|}(\zeta),
\end{align}
and the OP under the I.I.D. block-correlation channel approximation and CLT is given by
\begin{align}
	P_{\mbox{out}}\approx\frac{H\pi}{2U}\sum_{d=1}^U\sqrt{1-q_d^2}F_{{\bar{\gamma}^*}(y)}\left(\frac{\sqrt{\beta}-\Omega_2 \zeta}{\Omega_1}\right)f_{|\chi|}(\zeta).
\end{align}

\section{Numerical Results}  
In this section, we provide numerical results to verify the accuracy of the derived outage probability expressions. The RIS is configured to adjust $\theta_m$ to optimize the channel gain as outlined in Eq. \eqref{aa1} for the receiver. The simulation parameters are set as follows: $\epsilon_1=\epsilon_2=1$, $\epsilon_3=1/2$, and $W = 5$, $R=10$ bit/s/Hz, $\alpha=3.9$, $\sigma_k^2=\sigma_r^2=-40$ dBm. For the simulation setup, we use a two-dimensional Cartesian coordinate system. The BS is positioned  at (0, 0), the RIS at (40 m, 40 m), and the receiver at (100 m, 0 m), and $d_0=10$ m.

\begin{figure*}[t] 
		\vspace{-3mm}
		\begin{minipage}{0.32\linewidth}
			\centering
			\includegraphics[width=\textwidth]{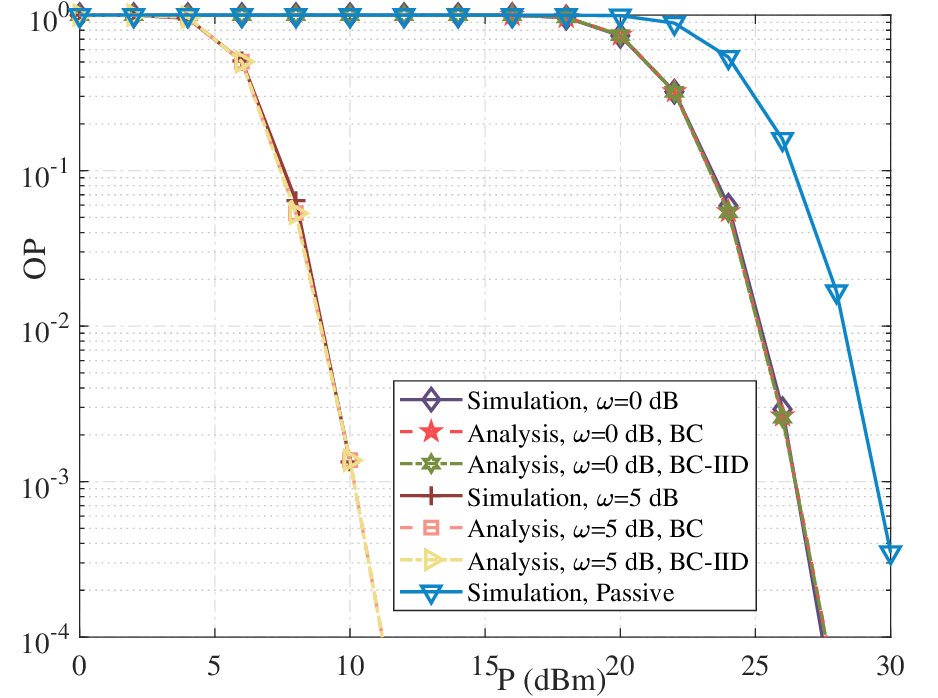}
			\caption{OP v.s. P.}\label{fig1}  
		\end{minipage}
		\hfill
	\begin{minipage}{0.32\linewidth} 
		\centering
		\includegraphics[width=0.98\textwidth]{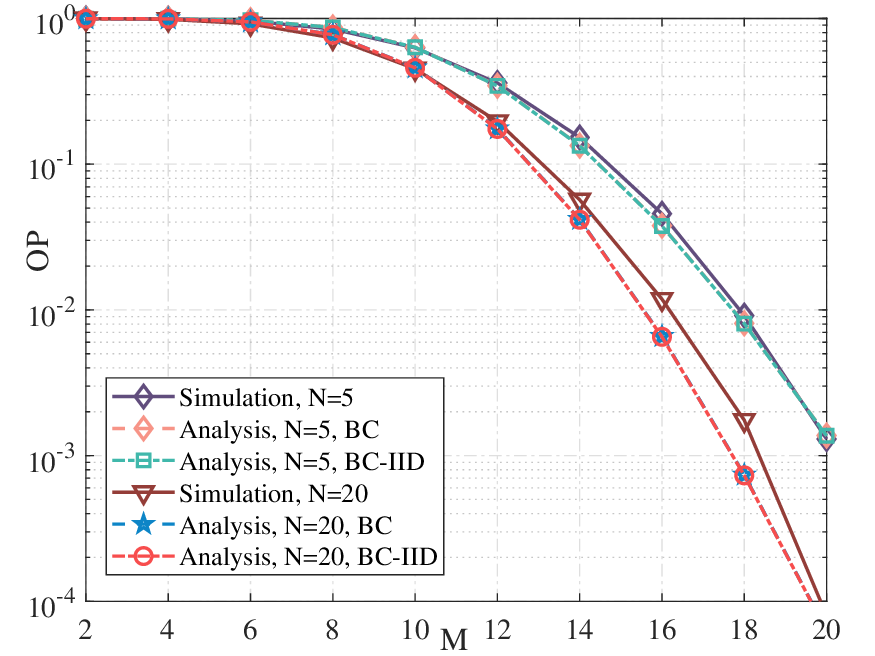} 
		\caption{OP v.s. M.}\label{fig2}\vspace{5mm}
	\end{minipage}
	\hfill
	\begin{minipage}{0.32\linewidth}\vspace{-3mm}
\centering
\includegraphics[width=0.98\textwidth]{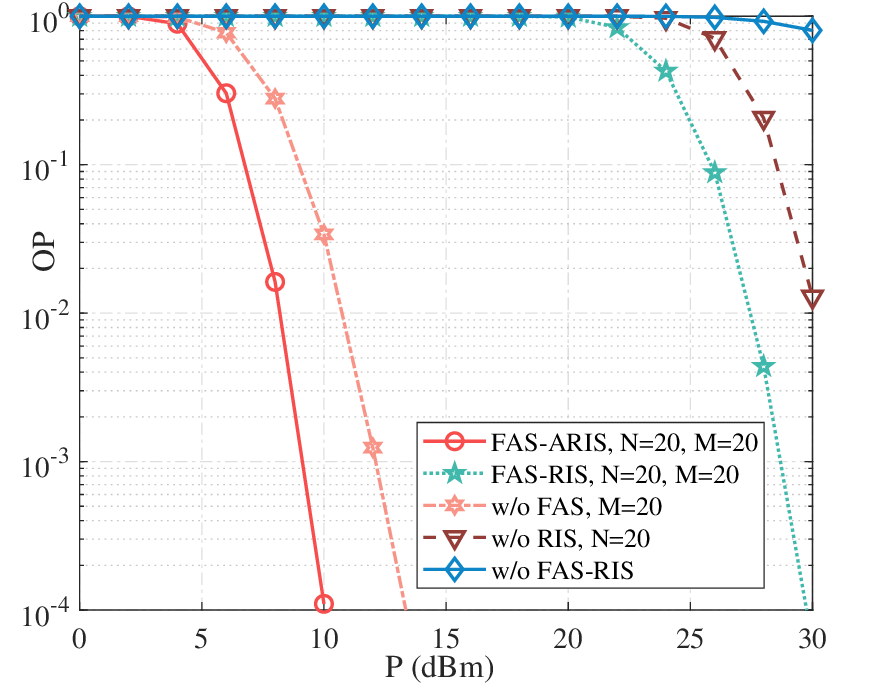}
\caption{OP v.s. P for different scenarios.}\label{fig3}
	\end{minipage} 
\end{figure*}

 In Fig.~\ref{fig1}, even when the amplification factor \(\omega = 0\) dB, the FAS-ARIS outperforms the FAS-RIS once the transmit power \(P\) exceeds 15 dBm, and this performance gap continues to widen as \(P\) increases. Additionally, with a small amplification factor of 5 dB, FAS-ARIS delivers strong communication performance even in low-power scenarios.  The theoretical results closely match the simulations, further validating the accuracy of the employed BC model for FAS-ARIS systems. Hence, we have the following remark.
\\ \textbf{Remark 1}:
 \textit{Under a low power budget, FAS-ARIS is the better choice compared to FAS-RIS, even with a relatively small amplification factor \(\omega = 5\) dB, as it effectively enhances communication performance.}

Fig.~\ref{fig2} depicts the relationship between the OP and $M$. We have set the $P=10$ dBm, $\omega=5$ dB, with $N = 5$ or $N = 20$.  The results  in Fig.~\ref{fig2}, demonstrate the consistency between the theoretical analysis and the simulation results for the OP, confirming the correctness of the derivation. Additionally, it is observed that increasing the number of ports \(N\) further improves the system performance, which highlights the benefits of the FAS.

Fig.~\ref{fig3} illustrates the relationship between the OP and $P$ for various configurations, comparing the performance of FAS-ARIS systems against scenarios with FAS-RIS without RIS, without FAS, and without FAS-ARIS. The curves clearly demonstrate that the FAS-ARIS configuration significantly outperforms all other setups, with ARIS playing a crucial role in improving outage performance. The effectiveness of FAS is also evident, as it provides additional performance gains when combined with ARIS, highlighting the importance of both technologies in enhancing communication performance. Hence, we have the following remark.
\\\textbf{Remark 2}: \textit{Our findings confirm that ARIS is crucial for system performance, as its absence leads to significant degradation. Integrating FAS with ARIS further enhances performance, making the FAS-ARIS system the most effective solution for maximizing communication. This win-win synergy provides valuable insights and guidance for the development of 6G networks.}

\section{Conclusion}
In this paper, we analyzed a FAS-ARIS communication systems. By employing the BC approximation model, we derived the approximate expressions for the OP. The simulation results reveal that FAS-ARIS system significantly outperforms other system.

\end{document}